\documentclass[a4paper]{jpconf}
\usepackage{graphicx}
\usepackage[colorlinks=true,linktocpage=true,
linkcolor=blue,citecolor=blue]{hyperref}
\usepackage[usenames,dvipsnames]{color}

\begin{document}
\title{Exact solutions of the (0+1)-dimensional kinetic equation in the relaxation time approximation}

\author{Ewa Maksymiuk}

\address{Institute od Physics, Jan Kochanowski University, PL-25460 Kielce, Poland}

\ead{MaksymiukEwa@gmail.com}

\begin{abstract}
We present exact solutions of the (0+1)-dimensional kinetic equation for a massive gas in the relaxation time approximation.  At first, we analyse the case of classical statistics and argue that the traditional second-order hydrodynamics misses the shear-bulk coupling. In the next step, we include Bose-Einstein and Fermi-Dirac statistics in the calculations and show that they are important for description of the effects connected with bulk viscosity.
\end{abstract}

\section{Introduction}
In this paper we present exact solutions of the (0+1)-dimensional kinetic (Boltzmann) equation for a massive gas in the relaxation time approximation~\cite{Florkowski:2014sfa}. Our results  describe a purely longitudinal boost-invariant expansion and may be useful for description of very early stages of relativistic heavy-ion collisions.  At first, we analyse the case of classical statistics and show that the traditional second-order hydrodynamics has problems to correctly reproduce the kinetic result. This discrepancy is due to the missing shear-bulk coupling in the standard second-order hydrodynamics. We further take into account the effects of quantum statistics~\cite{Florkowski:2014sda}. They turn out to be important for the bulk viscous pressure. 

The results presented here generalise several earlier results obtained for massless particles~\cite{Florkowski:2013lza,Florkowski:2013lya}.  We note that the exact solutions of the kinetic equation help us to select the right form of the kinetic coefficients~\cite{Florkowski:2013lza,Florkowski:2013lya} and the correct structure of the hydrodynamic equations, as has been demonstrated recently in Refs.~\cite{Florkowski:2013lza,Florkowski:2013lya,Florkowski:2014bba,Nopoush:2014pfa,Denicol:2014xca,Denicol:2014tha,Nopoush:2014qba}.

\section{The boost-invariant Boltzmann equation in the relaxation time approximation}

Our approach uses a simple form of the kinetic equation, namely
\begin{equation}
 p^\mu \partial_\mu  f =  C[f]\, ,\hspace{1.5cm} C[f] = - \frac{p_\alpha u^\alpha}{\tau_{\rm eq}} (f - f_{\rm eq}),
\label{kineq}
\end{equation}
where $f$ is the one-particle phase-space distribution function depending on the parton space-time coordinates $x$ and momentum $p$, and $C$ is the collision term written in the relaxation time approximation~\cite{1954PhRv...94..511B,Anderson:1974,Czyz:1986mr}.  The parameter $\tau_{\rm eq}$ is the relaxation time. In our present calculations we use the value $\tau_{\rm eq}=0.25\,\,\rm fm/c$. The boost-invariance implies that the kinetic equation (\ref{kineq}) can be rewritten in the form
\begin{equation}
\frac{\partial f}{\partial \tau} = 
\frac{f_{\rm eq}-f}{\tau_{\rm eq}},
\label{kineqs}
\end{equation}
where $\tau = \sqrt{t^2-z^2}$ is  the proper time. In addition, the function  $f$ may depend only on the three variables: $\tau$, $w$ and $p_T$, where $w=tp_L-zE$. The background equilibrium distribution function may be written as
\begin{equation}
f_{\rm eq}(\tau, w, p_T) =
\frac{2}{(2\pi)^3} \left\{ \exp\left[
\frac{\sqrt{w^2+ \left( m^2+p_T^2 \right) \tau^2}}{T(\tau) \tau} \,  \right] -\epsilon\right\}^{-1}.
\label{feq}
\end{equation}
In Eq.~(\ref{feq}) the parameter $\epsilon$ specifies the appropriate quantum statistics. With $\epsilon=+1,0,-1$ we include Bose-Einstein, Boltzmann or Fermi-Dirac statistics, respectively.

The first moment of the left-hand side of the kinetic equation describes divergence of the energy-momentum tensor that must be conserved, namely
\begin{equation}
T^{\mu\nu}(\tau) = g_0\int dP p^\mu p^\nu f(\tau,w,p_T),\hspace{1cm}\partial_{\mu}T^{\mu\nu}=0.
\label{Tmunu1}
\end{equation}
Here $g_0$ is the number of internal degrees of freedom and $dP$ is the momentum integration measure. Equation (\ref{Tmunu1}) is fulfilled if the energy densities calculated with the distribution functions $f$ and $f_{\rm eq}$ are
equal. This leads us to our main equation
\begin{eqnarray}
&&  
T^4(\tau)  \tilde{\cal H}_2\left[1,\frac{m}{T(\tau)}\right]
\label{LM2}  \\
&& 
 = D(\tau,\tau_0) \Lambda^4_0 \tilde{\cal H}_2\left[ \frac{\tau_0}{\tau \sqrt{1+\xi_0}},\frac{m}{\Lambda_0}\right]  + \int\limits_{\tau_0}^\tau 
\frac{d\tau^\prime}{\tau_{\rm eq}} D(\tau,\tau^\prime)
T^4(\tau^\prime) 
\tilde{\cal H}_2\left[ \frac{\tau^\prime}{\tau},\frac{m}{T(\tau^\prime)}\right]. \nonumber
\end{eqnarray}
This is an integral equation for the effective temperature $T(\tau)$ that can be solved using the iterative method \cite{Banerjee:1989by}. The function $\tilde{\cal H}_2(y,z)$ is defined in Ref.~\cite{Florkowski:2014sda}. Here we used the initial condition given by the Romatsche-Strickland form
\begin{eqnarray}
f_0(w,p_T) &=& \frac{1}{4\pi^3}
\left\{ \exp\left[
\frac{\sqrt{(1+\xi_0) w^2 + (m^2+p_T^2) \tau_0^2}}{\Lambda_0 \tau_0}\, \right] -\epsilon \right\}^{-1}.
\label{f0}
\end{eqnarray}
We note that the form of Eq.~(\ref{LM2}) is the same for the three different statistics. The differences are hidden in the implicit dependence of the functions $ f_{\rm eq}$, $f_0$, and $\tilde{\cal H}_2$ on the quantum statistics parameter $\epsilon$.

\section{Shear and bulk viscosities of a relativistic quantum massive gas}
To find the shear viscosity $\eta$ for the Bose--Einstein and Fermi--Dirac gases, we use the formula~\cite{Sasaki:2008fg}
\begin{eqnarray}
\eta(T)=\frac{2g_0\tau_{\rm eq}}{15T}\int\frac{d^3p}{(2\pi)^3}\frac{p^4}{E^2}\, f_{\rm eq}(1 +\epsilon f_{\rm eq}).
\label{eta_qs}
\end{eqnarray}
On the other hand, we determine the effective shear viscosity using the exact solution of the kinetic equation %
\begin{equation}
\eta_{\rm eff}(\tau) = \frac{1}{2} \, \tau \, \left[{\cal P}_T(\tau)-{\cal P}_L(\tau)\right]. 
\label{etakin}
\end{equation}
We treat the bulk viscosity $\zeta$ in the similar way as the shear viscosity. For a quantum massive gas, the formula for the bulk viscosity is the following~\cite{Sasaki:2008fg}
\begin{eqnarray}
\zeta(T)=\frac{2g_0\tau_{\rm eq}}{3T}\int\frac{d^3p}{(2\pi)^3}\frac{m^2}{E^2}
f_{\rm eq}(1 + \epsilon f_{\rm eq})\left(c_s^2E-\frac{p^2}{3E}  \right).
\label{zeta_qs}
\end{eqnarray}
To find the effective bulk viscosity, which is obtained from the exact solution we use the formula
\begin{equation}
\zeta_{\rm eff}(\tau) = -\frac{1}{3} \tau
\left[{\cal P}_L(\tau) + 2 {\cal P}_T(\tau)
- 3 {\cal P}_{\rm eq}(\tau) \right] .
\label{zetakin}
\end{equation}
The sound velocity appearing in Eq.~(\ref{zeta_qs}) is obtained from the formula $c_s^2(T) = \partial {\cal P}_{\rm eq}(T)/\partial {\cal E}_{\rm eq}(T)$. In the classical limit, $\epsilon \to 0$, the integrals (\ref{eta_qs}) and (\ref{zeta_qs}) become analytic. The appropriate formulas can be found in~Ref.~\cite{Florkowski:2014sfa}.

\section{Results}
\begin{figure}[t!]
\begin{center}
\begin{minipage}[b]{5.5cm}
\centering
\includegraphics[angle=0,width=1.\textwidth]{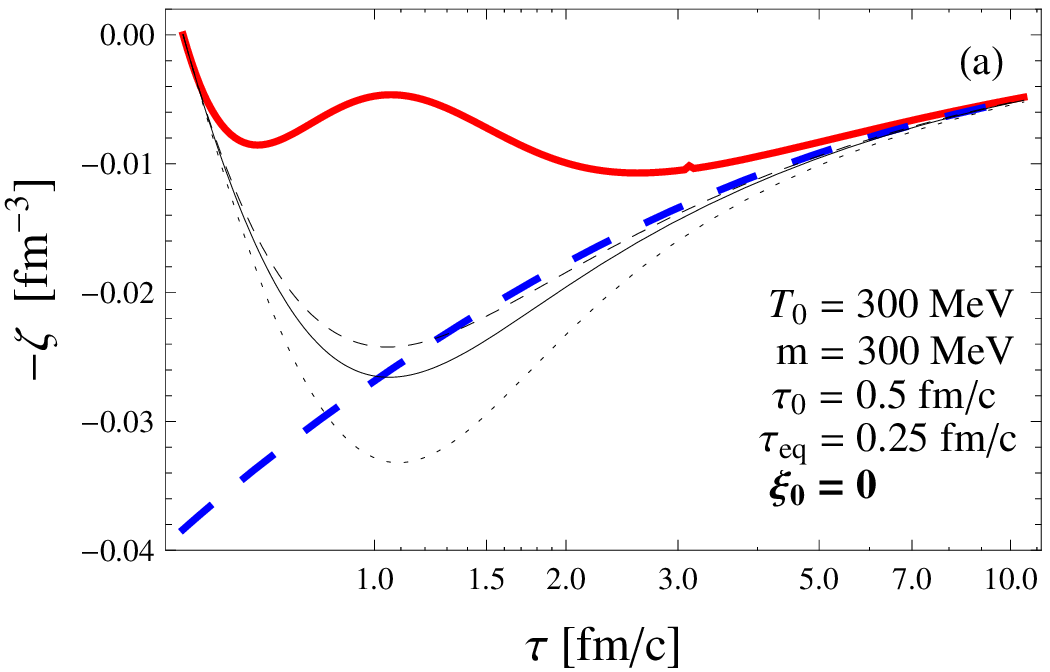}
\end{minipage}
\vspace{-0.1cm}
\begin{minipage}[b]{5.5cm}
\centering
\includegraphics[angle=0,width=0.95\textwidth]{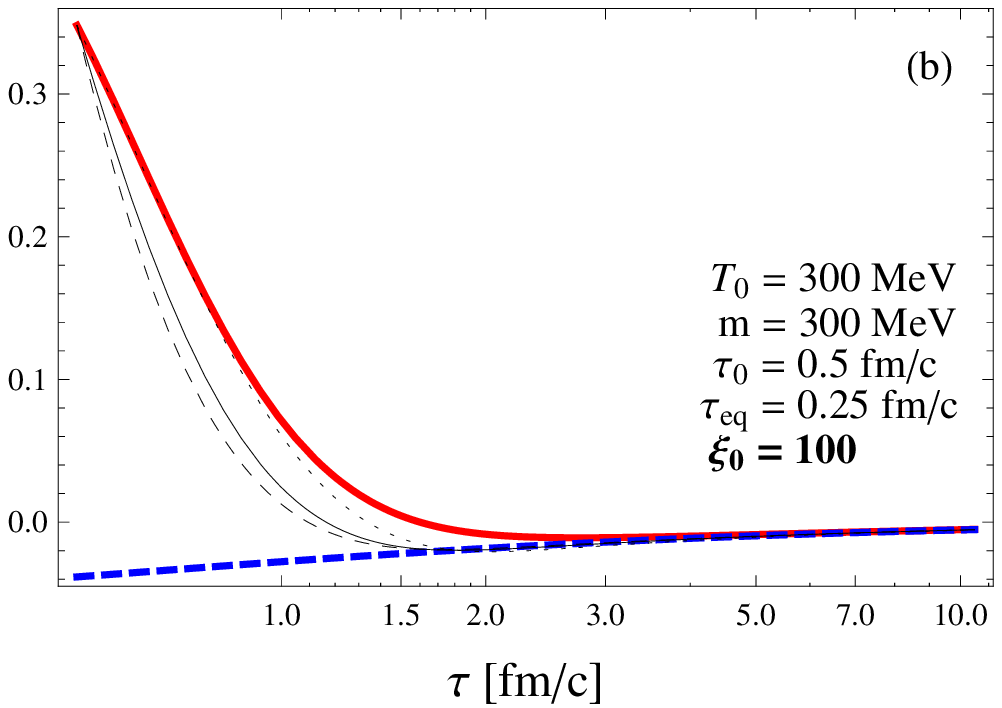}
\end{minipage}
\caption{Comparison of the bulk viscosity for Boltzmann statistics. The panel (a) describes a system which is initially isotropic, while the panel (b) describes a system which is initially highly oblate. Red solid lines represent the effective bulk viscosity obtained from the kinetic theory, blue-dashed lines describe the bulk viscosity given by Eq. (\ref{zeta_qs}), finally,  black lines show our results obtained for three versions of the second--order hydrodynamics \cite{Florkowski:2014sfa}.}
\label{fig:Bulk_Boltzmann}
\end{center}
\end{figure}
Results shown in Fig.~\ref{fig:Bulk_Boltzmann} represent the time evolution of the bulk viscosity calculated directly from the kinetic theory and compared with three formulations of the second-order hydrodynamics~\cite{Florkowski:2014sfa}. In the case of an initially isotropic system we can see that the simplest formulation of the second-order hydrodynamics (dotted line) gives the worst agreement with the kinetic theory. On the other hand,  this formulation gives the best agreement for an initially highly oblate system. The problems illustrated in  Fig.~\ref{fig:Bulk_Boltzmann} helped to identify the importance of the shear-bulk couplings \cite{Denicol:2014vaa,Denicol:2014mca,Jaiswal:2014isa}  in the hydrodynamic approach. To have a good agreement it is necessary to use equations where the bulk and shear viscosities are correlated~\cite{Denicol:2014mca,Jaiswal:2014isa}. The proper description of the bulk pressure is important as it may affect different physical observables studied in relativistic heavy-ion collisions~ \cite{Noronha-Hostler:2013gga,Noronha-Hostler:2013hsa,Rose:2014fba}. 

\begin{figure}[t!]
\begin{center}
\begin{minipage}[b]{5.5cm}
\centering
\includegraphics[angle=0,width=1\textwidth]{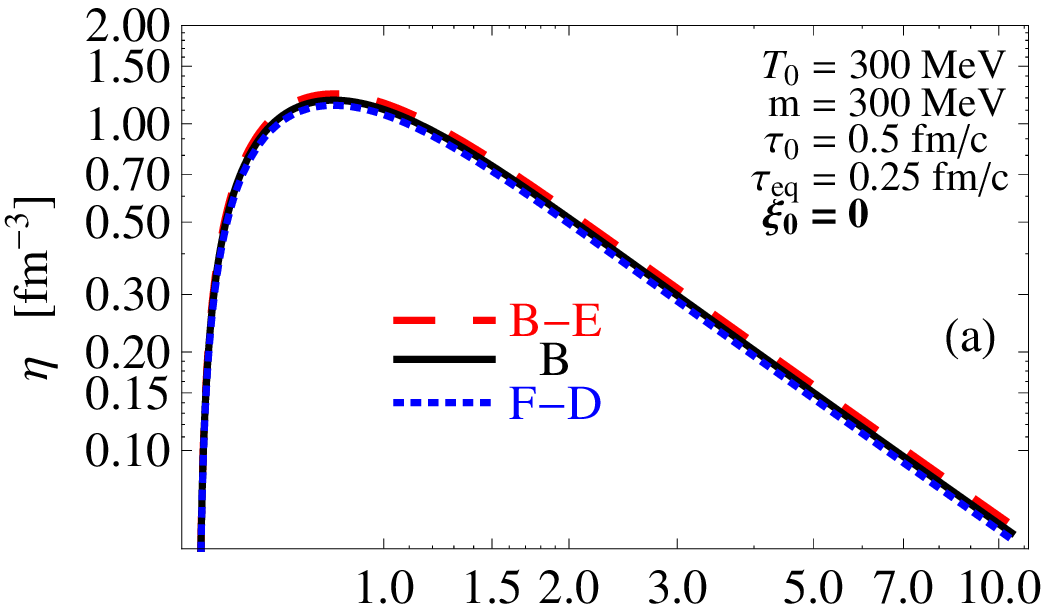}
\end{minipage}
\vspace{-0.1cm}
\begin{minipage}[b]{5.5cm}
\centering
\includegraphics[angle=0,width=0.96\textwidth]{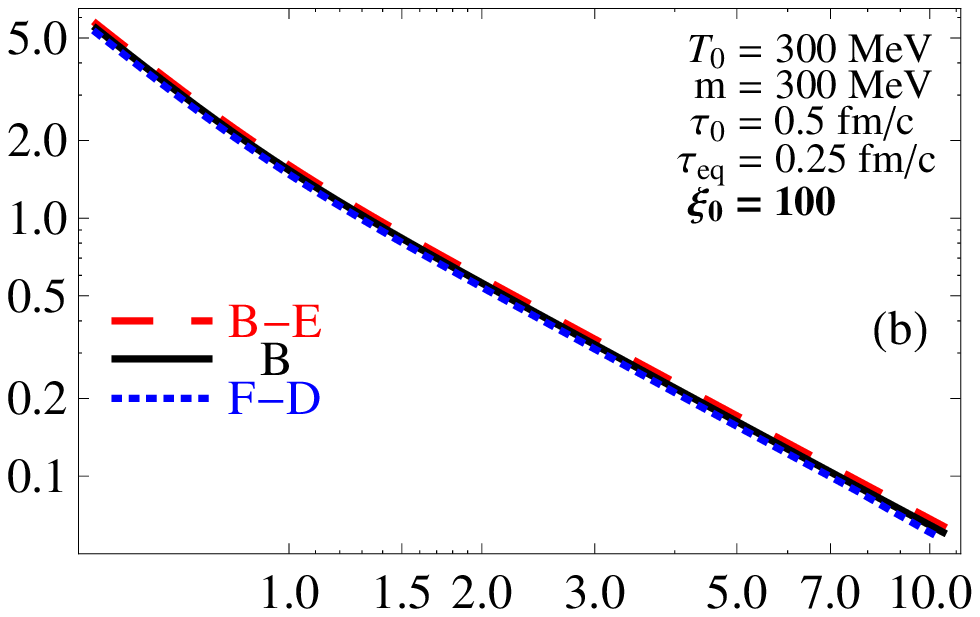}
\end{minipage}
\\
\begin{minipage}[b]{5.5cm}
\centering
\includegraphics[angle=0,width=1\textwidth]{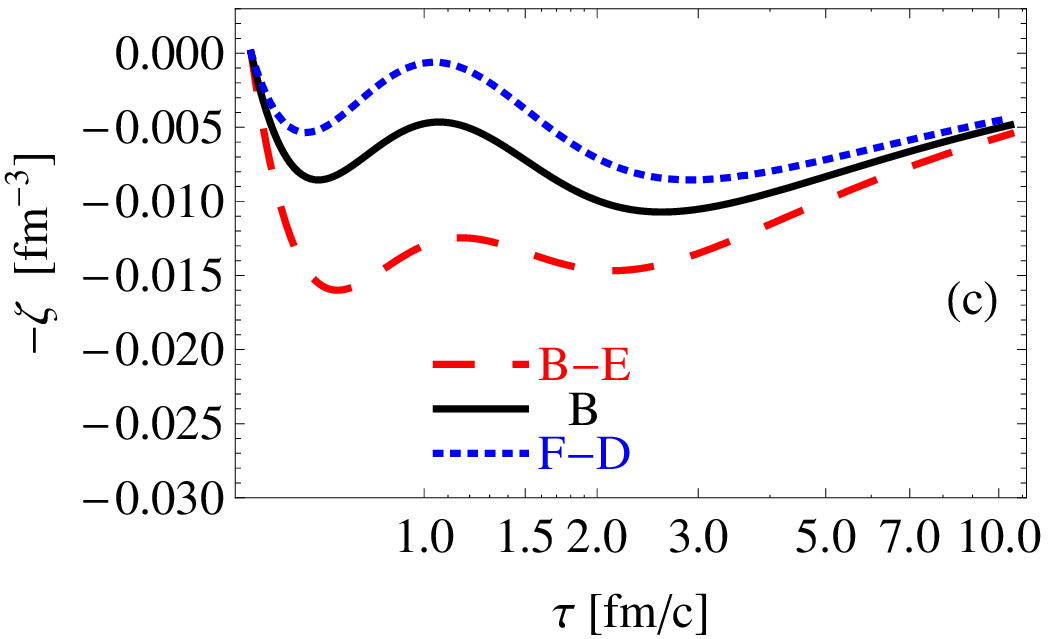}
\end{minipage}
\begin{minipage}[b]{5.5cm}
\centering
\includegraphics[angle=0,width=0.95\textwidth]{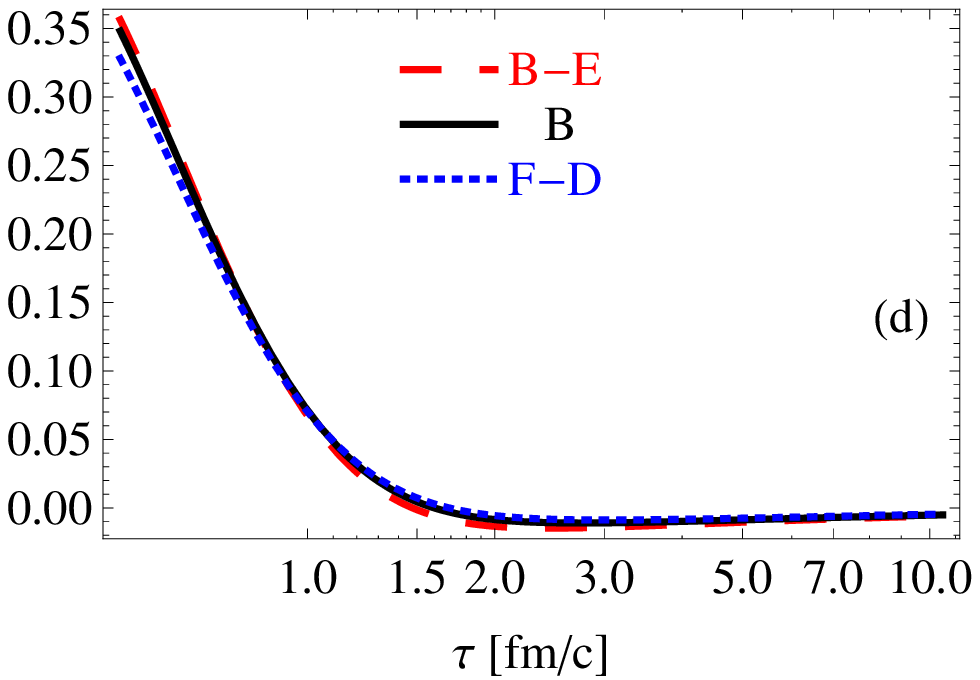}
\end{minipage}
\caption{Comparison of exact solutions for different quantum statistics. The values of the parameters are displayed in the panels. Panels (a) and (c) differ from (b) and (d) by the value of the initial anisotropy parameter. }
\label{fig:Quantum_statistics_Comparison}
\end{center}
\end{figure}

To show the effects of  quantum statistics on the evolution of matter we calculated the shear viscosity using Eqs.~(\ref{eta_qs}) and ~(\ref{etakin}) and the bulk viscosity using Eqs.~(\ref{zeta_qs}) and~(\ref{zetakin}). In Fig.~\ref{fig:Quantum_statistics_Comparison} we present our results. In the case of the shear viscosity, we find only small differences between the results obtained for Bose-Einstein, Boltzmann, and Fermi-Dirac statistics. This is in contrast to the bulk-viscosity case, where we find important differences,  especially, for the initially isotropic systems.

\section{Conclusions}
We have constructed the exact solutions of the Boltzmann equation using analytical and numerical methods. This allowed us to find the effective bulk and shear viscosities of the (0+1)--dimensional system and to compare them with the analytic formulas appearing in the literature. We have shown that standard equations of the second-order hydrodynamics do not work properly --- although the quantum statistics effects are not essential for the shear viscosity they become quite important for the correct description of the bulk viscous effects.

\ack
I would like to thank my supervisor Wojciech Florkowski for valuable advices and interesting discussions. 
This work was supported by Polish National Science Center grant No. DEC-2012/06/A/ST2/00390.

\section*{References}

\end{document}